# Experimental demonstration of a two-dimensional phonon cavity in the quantum regime

A. N. Bolgar[1]*, J. I. Zotova[1], D. D. Kirichenko[1], I. S. Besedin[2], A. V. Semenov[1,3], R. S. Shaikhaidarov[1,4] and O. V. Astafiev[1,4,5]

[1]Moscow Institute of Physics and Technology, Institutskiy per. 9, Dolgoprudny, Russia
[2]National University for Science and Technology (MISiS), Leninskiy pr. 4, Moscow 119049, Russia
[3]Moscow State Pedagogical University, Malaya Pirogovskaya str. 1/1, Moscow 119435, Russia
[4]Royal Holloway University of London, Egham Surrey TW20 0EX, United Kingdom
[5]National Physical Laboratory, Teddington, TW11 0LW, United Kingdom

The quantum regime in acoustic systems is a focus of recent fundamental research in the new field of Quantum Acoustodynamics (QAD). Systems based on surface acoustic waves having an advantage of easy integration in two-dimensions are particularly promising for the demonstration of novel effects in QAD and development of novel devices of quantum acousto-electronics. We demonstrate the vacuum mode of the surface acoustic wave resonator by coupling it to a superconducting artificial atom. The artificial atom is implemented into the resonator formed by two Brag mirrors. The results are consistent with expectations supported by the system model and our calculations. This work opens the way to map analogues of quantum optical effects into acoustic systems.

Quantum Acoustodynamics is a new direction of quantum mechanics studying interaction of acoustic waves and phonons with quantum systems *(1-6)* and particularly with artificial quantum systems *(7-9)*. Recently quantum acoustics has been focused on establishing the quantum regime in phonon systems *(10-13)*. The coupling to the vacuum mechanical mode bulk acoustic resonators has been demonstrated in Ref. 1. Although, this was a very important achievement, such an approach is hard for further development because of difficulties in implementation of the bulk resonators and their integration in two dimensional circuits. As it has been recently demonstrated in Ref. 3, these problems are to overcome by utilizing surface acoustic waves (SAW). Here, we demonstrate the interaction of an artificial atom with a quantized vacuum mode of a two-dimensional SAW resonator on a quartz crystal, observed as an energy level splitting with a vacuum resonator mode. Further development will result in realisation of a series of analogues of different quantum optical effects in two-dimensions and can help create compact elements for quantum informatics.

Surface acoustic waves in piezoelectrics were used for a long time in compact electronic elements operating in a megahertz range such as RF filters, resonators, delay lines etc. *(14-18)*. One of the main advantages of planar SAW devices is their small size conditioned by a slow speed of sound compared to electromagnetic waves and therefore up to five orders of magnitude shorter wavelengths for the same frequencies. It has been recently shown that superconducting artificial atoms, successfully exploited for coherent control of photons *(19)* and demonstration of quantum optics with single quantum systems *(20)*, can also be used for control of single phonons and more generally for quantum acoustodynamics *(2,3,21)*. The next important milestone must be a demonstration of an interaction of the artificial atom with quantised resonator modes. Technologically, reaching the quantum regime is a challenging task due to submicron-scale wavelengths in the microwave range, which requires state-of-the-art nano-fabrication methods.

We consider a SAW resonator interacting with a superconducting two-level system described by the standard Jaynes-Cummings Hamiltonian *(22)*

$$H = \frac{\hbar \omega_a}{2}\sigma_z + \hbar \omega_r b^\dagger b + \hbar g\left(b^\dagger \sigma^- + b\sigma^+\right), \quad (1)$$

where $\sigma_z$ is the Pauli matrix, $\sigma^+$ ($\sigma^-$) is the creation (annihilation) operator of the two-level system excited state, $b^\dagger$ ($b$) is the creation (annihilation) phonon operator. The first term represents the two-level atom with energy splitting $\hbar \omega_a$, the second describes the SAW resonator with resonant frequency $\omega_r$ and the third is resonator-atom interaction with coupling strength $\hbar g$. The same Hamiltonian, only with photon ladder operators instead of phonon, is commonly used to describe interaction of an atom with quantized electromagnetic fields.

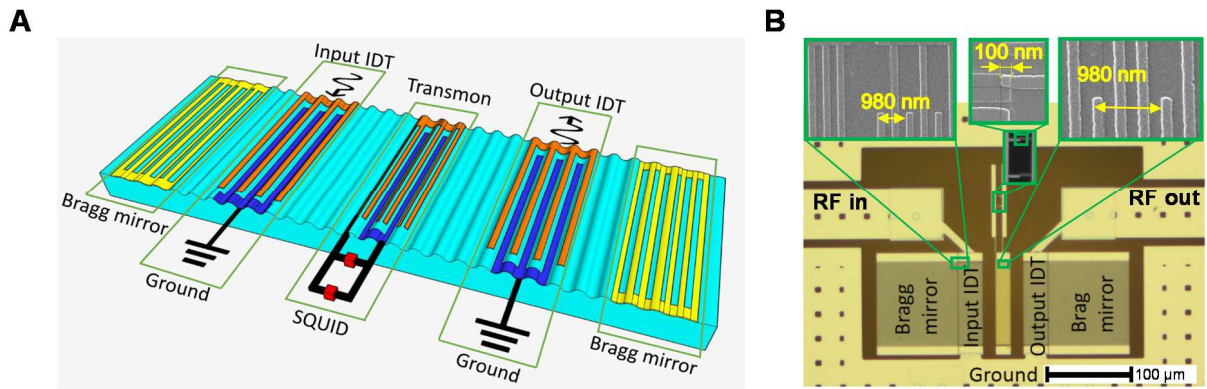

**Fig. 1.** The device. **(A)** A schematic 3D representation of the sample. Two identical IDT ports and the transmon qubit are located inside the SAW cavity. The SAW resonator is formed by two Bragg gratings, consisting of 200 periodic stripe electrodes each, with a period *p*, where *p* = 980 nm. Distance between the gratings is 225 × *p*/2 = 110.7 μm. IDT ports consist of 29 periodic identical cells with a period *p*. Each cell consists of 2 electrodes, each connected to bars at opposite sides. The distance between the IDT port and the adjacent grating is $d_1$ = (1 + 1/8) *p*. The transmon qubit is located between the IDT ports and it consists of a SQUID shunted by an IDT capacitance. The qubit IDT has 18 cells with 3 electrodes per period *p*, which minimizes its mechanical reflection of acoustic waves, in order to suppress parasitic resonances this element would otherwise cause. The length of all electrodes in our device is W = 100 μm. The width of the electrodes of the gratings and IDT ports is *p*/4, and the width of the qubit IDT electrodes is *p*/6. **(B)** Optical micrographs of the sample with insets showing the zoomed stripe structures and the Josephson junction, obtained with a scanning electron microscope (SEM). All IDTs and the SQUID are fabricated on a ST-X cut of a quartz crystal substrate by means of electron-beam lithography. The ground plane and coplanar lines are fabricated with the use of optical lithography. All metal structures are formed by aluminium electron-beam evaporation.

Our device, shown in Fig. 1, is fabricated on a quartz substrate. It consists of an acoustic resonator with a superconducting artificial atom inside. The resonator is a 2D Fabry-Perot cavity, formed by two Bragg gratings *(23,24)*. To excite and to detect the SAW we add two identical interdigitated transducer ports (IDTs) inside the resonator. The IDT converts applied ac-voltage into an acoustic wave and it is formed by a periodic array of alternating electrodes with a fixed pattern per period. We use two distinct IDT patterns with 2 and 3 electrodes per period, as shown in Fig. 1.

A tuneable two-level artificial atom consists of a SQUID shunted by an IDT structure, playing the role of both a qubit capacitance and a coupler to SAWs in the resonator. The qubit IDT has the same period as IDT ports and mirrors and its electrodes are positioned at the expected antinodes of standing acoustic wave in the resonator. The calculated qubit capacitance is $C_\Sigma \approx 90$ fF, which corresponds to the charging energy $E_C/h = 0.21$ GHz, where $E_C = e^2/2C_\Sigma$. The SQUID consists of two Josephson junctions with maximal Josephson energy $E_{J0}/h = 17$ GHz. The energy of the qubit is controlled by a magnetic field of a surrounding solenoid, which tunes the effective Josephson energy $E_J$ of the SQUID. The ratio $E_{J0}/E_C = 80$ defines the transmon regime and the ground-to-first excited state transition energy $E_{01} \approx \sqrt{8E_J E_C} - E_C$.

The SAW propagation speed at low temperatures in quartz is $v \approx 3.16$ km/s *(21)*. The periodicity of IDT stripes is $p = 980$ nm and that of the Bragg mirrors is $p/2$, which defines an optimal SAW wavelength and frequency: $\lambda = p$, $\omega_0/2\pi \approx 3.2$ GHz. Each element of our sample has a finite frequency bandwidth determined by its geometry. The calculated frequency characteristics of different elements are plotted in Fig. 2A. The resulting bandwidth of the resonator is limited by the Bragg mirrors to about 33 MHz.

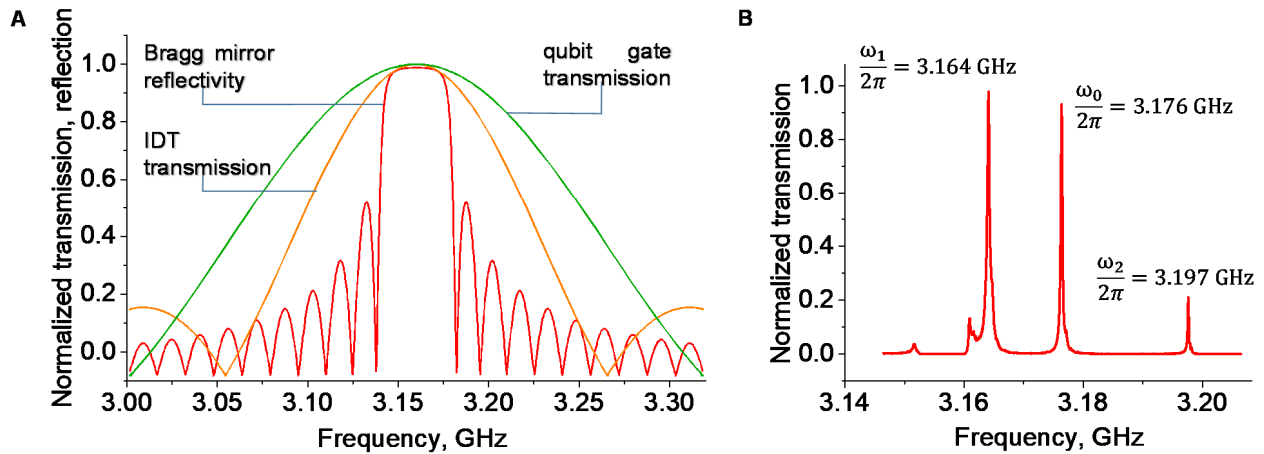

**Fig. 2.** Frequency characteristics. **(A)** Calculated frequency characteristics of different circuit elements. The Bragg mirror reflectivity (red line), has the width of the main maximum $\Delta F_m = 33$ MHz. Green and yellow curves correspond to the simulated frequency dependence of the absolute values of transmission amplitudes for a signal, applied to the input IDT (yellow) or qubit IDT (green). The corresponding bandwidths are $\Delta F_{IDT} = 95$ MHz and $\Delta F_q = 143$ MHz. **(B)** Measured transmission through a SAW resonator. The central peak corresponds to the SAW mode, interacting with the qubit, since its IDT electrodes are located in acoustic field antinodes for this mode.

All measurements described below are performed in a dilution refrigerator at the base temperature of 20 mK. To measure the acoustic response, we implement the same method and the measurement circuit as the ones used in quantum optics experiments with the superconducting artificial atoms described for example in ref. *(19,22)*. The electromagnetic microwaves are transmitted from a vector network analyser (VNA) through coaxial cables and then through an on-chip coplanar line to the input IDT port, where they are converted into SAW in the resonator. The standing SAWs are converted into electromagnetic waves by the output IDT port. Then the signal is amplified by cryogenic amplifiers and measured by the VNA.

We first perform a measurement of transmission amplitude through the resonator, which is shown on Fig. 2B. In this measurement the qubit is not tuned into resonator frequency. We find three resonances with frequencies $\omega_1/2\pi = 3.164$ GHz, $\omega_0/2\pi = 3.176$ GHz and $\omega_2/2\pi = 3.197$ GHz. The full-width-at-half-maximum (FWHM) of a power peak at $\omega_0$ is $\Delta\omega/2\pi = 0.332$ MHz.

We monitor the complex transmission amplitude $t$ through the cavity at frequency $\omega_0$ as a function of magnetic field applying a weak driving power to keep the effective phonon number in the resonator less than one. The transmission peak exhibits periodic anticrossings with the period corresponding to a flux quantum through the SQUID loop. The typical anticrossing is shown in Fig. 3. The anticrossing is a signature of interaction of the two-level system with the cavity and it disappears with an increase of the driving amplitude.

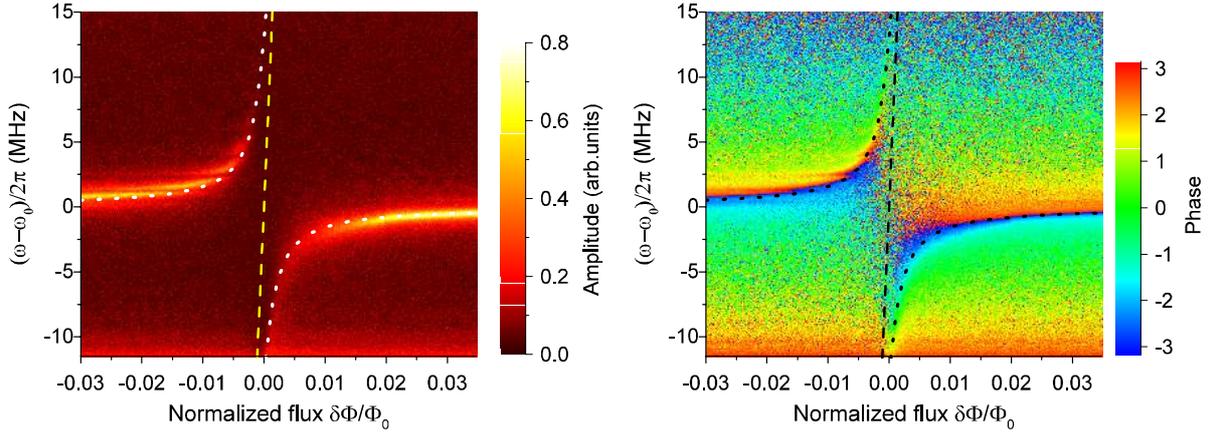

**Fig. 3.** Interaction of the qubit with the acoustic resonator. Left and right panels represent amplitude (|*t*|) and phase (arg(*t*) of the transmission coefficient *t* through the SAW resonator in vicinity of the resonator frequency $\omega_0$. The anticrossing demonstrates the interaction between the qubit and the resonator, when they are in resonance. The dashed line is the expected qubit energy without the interaction. The dotted line is the calculated energy splitting according to equation (2) with $g/2\pi$ = 13 MHz.

To find an energy splitting of this anticrossing for the lowest excitation we use the following formula *(25)*:

$$E_{\pm} = \frac{\hbar(\omega_0 + \omega_q) \pm \hbar\sqrt{(\omega_0 - \omega_q)^2 + 4g^2}}{2}, \qquad (2)$$

found from diagonalization of the Hamiltonian from equation (1). Using this function, we acquire the coupling constant $g/2\pi$ = 13 MHz from the fit of our anticrossing spectrum, as shown on Fig. 3a-b. This means that total splitting is $2g/2\pi$ = 26 MHz, which tightly fits in the bandwidth of the resonator $\Delta F_m$ = 33 MHz. Note also that the splitting does not occur at the adjacent resonator modes. The reason is that in the contrast with the central mode ($\omega_0$), these modes ($\omega_1$, $\omega_2$) correspond to standing waves in the resonator with nodes at the qubit gate electrode locations.

Next, using the interaction of the qubit with the cavity, we apply a method of dispersive readout known from the circuit QED *(22)* to characterize the qubit energies. By monitoring the transmission amplitude through the cavity at $\omega_0$, we sweep the second probe tone $\omega_p$ and find a pattern with one well distinguished qubit resonance line, corresponding to qubit transition energy $E_{10}$ (Fig. 4). We also see a signature of a $E_{12}$ transition as well as some higher order transitions. The pattern is periodic with the applied field with the period of flux quantum $\Phi_0$ through the SQUID. Measured qubit transition energies are in good agreement with our expectations. Note that the required qubit excitation in a wide frequency range takes place due to a weak but finite electric coupling of the input IDT to the qubit. From our capacitance simulations, we find the effective qubit gate capacitance to be $C_g \approx 0.1$ fF.

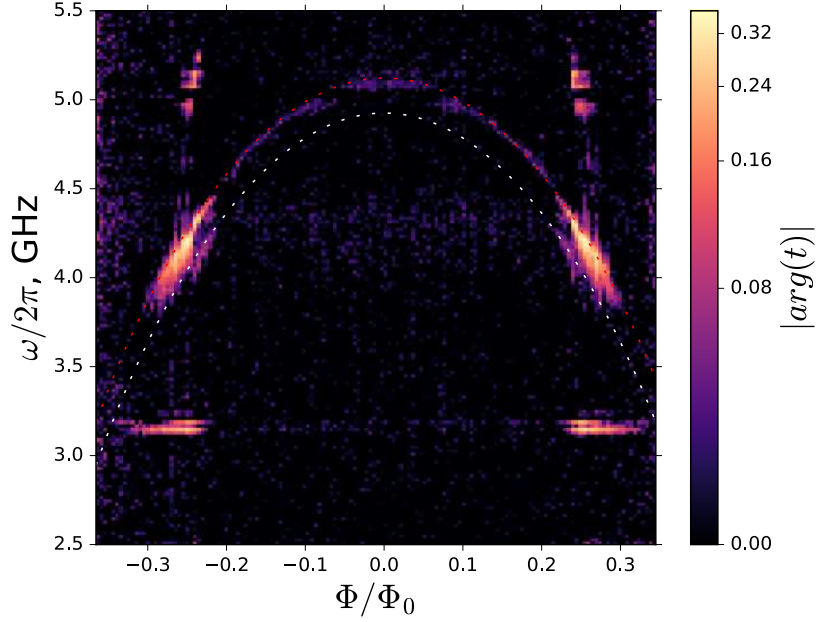

**Fig. 4.** Two tone spectroscopy. The phase shift of the first tone signal transmission is shown by colour. The vertical axis corresponds to the frequency sweep of the second tone. The horizontal axis corresponds to the magnetic flux through the SQUID. The dotted lines are the fitting curves, obtained from the qubit Hamiltonian (equation (1)) eigenstates calculation: $E_{01}$ (red), $E_{12}$ (white). The acoustic resonance is also seen at 3.176 GHz. There are also signatures of some other lines corresponding to higher order processes. From $E_{01}$ fits we obtain qubit's charging energy $E_C$ = 0.21 GHz and maximum Josephson energy $E_{J0}$ = 17.4 GHz, which is in good agreement with our expectations.

We also extract the relaxation rate of the qubit to be $\Gamma_1/2\pi \approx 10$ MHz (assuming no pure dephasing usually valid for tarnsmons) from an intrinsic width of the spectral line of the qubit by measuring it in the dispersive regime slightly away from the resonant point in the low drive limit. In the anticrossing, the width of the spectral lines close to the resonance defined by the qubit relaxation ($\Gamma_1/2$) is found roughly to be 5 MHz.

Now we independently estimate the coupling $g$ of the qubit to the resonator according to $\hbar g = \varsigma e V_0^2$, where $\varsigma = (E_J/32E_C)^{1/4} \approx 1$, $e$ is the elementary charge, $V_0$ is the amplitude of zero-point fluctuations induced in its turn by zero-point displacement (mechanical) fluctuations $U_0$ of the SAW mode. We find that $V_0 = e_{pz}/\varepsilon\, U_0$ and $e_{pz}/\varepsilon = 2\times 10^9$ V/m is the piezoelectric module normalized to quartz dielectric constant $\varepsilon$. Using the relation $U_0 = \sqrt{\hbar/2\rho A_c v}$, where the quartz density $\rho$ = 2647 kg/m$^3$ and the effective cavity area $A_c \approx 110$ μm × 140 μm = 1.5×10$^{-8}$ m$^2$, we estimate the coupling coefficient to be: $g/2\pi \approx 10$ MHz. This value is close to the experimentally measured one.

Finally, in order to better understand the exact mechanism of our system excitation, we consider two driving terms in the full Hamiltonian: (i) the drive of the acoustic cavity via IDT: $H_{ac} = \hbar\Omega_{ac}(b^\dagger + b)\cos\omega t$ and (ii) the electric drive of the qubit: $H_{el} = \hbar\Omega_{el}(\sigma^+ + \sigma^-)\cos\omega t$, where $\Omega_{ac}$ and $\Omega_{el}$ are acoustic and electric driving amplitudes respectively. The acoustic driving amplitude at the resonance can be found as $\hbar\Omega_{ac} = \mu_{ac}V$, where $\mu_{ac} = C_{IDT}V_0$ ($\approx 0.025\,e$) is the coupling between the voltage $V$ applied to the input IDT and the resonator driving amplitude $\Omega_{ac}$. The electric coupling between the IDT and the qubit can be expressed as $\hbar\Omega_{el} = \mu_{el}V$, where $\mu_{el} = C_g V_q$ and $V_q = \zeta 2e/C_\Sigma$ is the potential induced in the transmon qubit due to single Cooper pair transition.

Substituting the numbers, we find $\mu_{el} = 2e\, C_g / C_\Sigma \approx 0.002\, e$. This means that at the resonance the electric coupling is weaker than the acoustic one ($\mu_{ac}/\mu_{el} \approx 13$).

In conclusion, we have experimentally demonstrated interaction between an artificial atom and a SAW resonator. The result is an important milestone for the future realization of quantum acoustics effects dual to quantum optics *(27)* and also to build compact devices of quantum informatics *(26)*.

28. We would like to express our deep gratitude to Kirill Shulga and Nikolay Abramov for their contribution to the low temperature measurements. We also thank Aleksey Dmitriev and Alexander Korenkov for valuable discussions and their help in our experimental data analysis. We acknowledge Russian Science Foundation (grant N 16-12-00070) for supporting the work. This work was performed using technological equipment of MIPT Shared Facilities Center and with financial support from the Ministry of Education and Science of the Russian Federation (Grant No. RFMEFI59417X0014).